\documentclass[aps,prd,showpacs,floatfix,nofootinbib,twocolumn,10pt]{revtex4}

\usepackage{color,amsmath,amssymb,graphicx,latexsym,subfigure}
\usepackage{threeparttable,multirow,txfonts}

\begin{document}

\title{Implications on cosmic ray injection and propagation parameters
from Voyager/ACE/AMS-02 nucleus data}

\author{Qiang Yuan$^{a,b,c}$\footnote{yuanq@pmo.ac.cn}}

\affiliation{
$^a$Key Laboratory of Dark Matter and Space Astronomy, Purple Mountain
Observatory, Chinese Academy of Sciences, Nanjing 210008, P.R.China \\
$^b$School of Astronomy and Space Science, University of Science and
Technology of China, Hefei 230026, P.R.China\\
$^c$Center for High Energy Physics, Peking University, Beijing 100871,
P.R.China
}

\begin{abstract}

We study the propagation and injection models of cosmic rays using the
latest measurements of the Boron-to-Carbon ratio and fluxes of protons,
Helium, Carbon, and Oxygen nuclei by the Alpha Magnetic Spectrometer and
the Advanced Composition Explorer at top of the Earth, and the Voyager
spacecraft outside the heliosphere. The ACE data during the same time
interval of the AMS-02 data are extracted to minimize the complexity of
the solar modulation effect. We find that the cosmic ray nucleus data
favor a modified version of the diffusion-reacceleration scenario of the
propagation. The diffusion coefficient is, however, required to increase
moderately with decreasing rigidity at low energies, which has interesting
implications on the particle and plasma interaction in the Milky Way.
We further find that the low rigidity ($<$ a few GV) injection spectra
are different for different compositions. The injection spectra are
softer for lighter nuclei. These results are expected to be helpful
in understanding the acceleration process of cosmic rays.

\end{abstract}

\date{\today}

\pacs{95.85.Ry,96.50.S-,98.38.−j,94.20.wc}

\maketitle

\section{Introduction}

Precise measurements of the energy spectra and composition of cosmic
rays (CRs) provide us very important insights in understanding such
fundamental questions as the origin and propagation of CRs.
In particular, a good understanding of the propagation model and
injection parameters of background CRs is crucial for the indirect
detection of dark matter particles. The propagation parameters are
important for the calculation of dark matter signals, and the
injection parameters of primary CRs are relevant to the prediction
of background fluxes (e.g., positrons and antiprotons).
The secondary-to-primary nucleus ratio, such as the Boron-to-Carbon
ratio (B/C), is usually used to infer/constrain the propagation
models and parameters (e.g., \cite{1990ApJ...349..625S,
1991ApJ...374..356M,2001ApJ...555..585M,2011ApJ...729..106T,
2015JCAP...09..049J,2016ApJ...824...16J,2016PhRvD..94l3007F,
2016PhRvD..94l3019K}. With the new measurement of the B/C ratio by
e.g., the Alpha Magnetic Spectrometer (AMS-02) \cite{2016PhRvL.117w1102A},
significantly improved constraints on the model parameters can be obtained
\cite{2017PhRvD..95h3007Y,2018PhRvD..97b3015N,2018JCAP...01..055R}.

Just recently, the AMS-02 collaboration reported new measurements
of the primary (Helium, Carbon, Oxygen) and secondary (Lithium,
Beryllium, Boron) fluxes of CR nuclei up to rigidities of several
TV \cite{2017PhRvL.119y1101A,2018PhRvL.120b1101A}. These results
show very interestingly that these primary (or secondary) nuclei
share almost identical spectra with each other. More importantly,
there are spectral breaks of both primary and secondary nuclei at
a few hundred GV, and the spectral indices of secondary nuclei
harden by $\sim 0.13$ more than that of primary nuclei. Such a result
suggests a propagation origin of the spectral hardenings of CR nuclei
\cite{2012PhRvL.109f1101B,2012ApJ...752L..13T,2016ApJ...819...54G,
2017PhRvL.119x1101G,2018PhRvD..97f3008G,2018arXiv180203602L}.

However, the apparent similarity among the top-of-atmosphere (TOA) fluxes
of CRs does not reveal the acceleration properties of CRs directly,
due to the complicated intermediate steps such as the propagation in
the Galaxy and the solar modulation. With proper modeling of such
effects, it is possible to derive both the injection and propagation
parameters of CRs simultaneously. The difference of the intrinsic
injection spectrum among different species can then be used to
study the acceleration process of CRs.

In this work, we use the new results about the Carbon fluxes and B/C
ratio reported by AMS-02 to study the injection and propagation of CRs.
Compared with Ref. \cite{2017PhRvD..95h3007Y} in which the proton fluxes
and B/C ratio observed by AMS-02 and/or PAMELA were used, the use of
Carbon fluxes and B/C ratio can avoid the potential difference of the
injection spectra between protons and Carbon nuclei. We will also
include the Voyager data out of the solar system
\cite{2013Sci...341..150S,2016ApJ...831...18C} in this study, which
is expected to break the degeneracy between the injection spectra and
the solar modulation. The measurements of fluxes of CR nuclei by the
Cosmic Ray Isotope Spectrometer (CRIS) on the Advanced Composition
Explorer (ACE) spacecraft are included too. Finally, all the data
measured at TOA used in this work were taken at the same time, which
again can minimize the complexity of the solar modulation.

We use the {\tt CosRayMC} code we developed in past years
\cite{2010PhRvD..81b3516L,2012PhRvD..85d3507L}, which embeded the
numerical CR propagation code {\tt GALPROP}
\cite{1998ApJ...509..212S,1998ApJ...493..694M} into the Markov Chain
Monte Carlo sampler adapted from {\tt CosmoMC} \cite{2002PhRvD..66j3511L},
to do the global fitting of the model parameters. In Sec. II, we briefly
describe the CR injection and propagation model settings. In Sec. III we
describe the ACE-CRIS data. Sec. IV presents the fitting results.
Some discussions about the results are given in Sec. IV.
Finally we conclude in Sec. V.

\section{Injection and propagation of CRs}

The injection spectrum of CR nuclei is assumed to be a broken power-law
form of rigidity
\begin{equation}
q(R)\propto\left\{
\begin{array}{ll}
\left( R / R_{\mathrm{br}}\right)^{-\nu_1}, & R < R_{\mathrm{br}} \\
\left( R / R_{\mathrm{br}} \right)^{-\nu_2}, &  R \ge R_{\mathrm{br}}
\end{array}.\right.
\label{injection}
\end{equation}
The parameters $\nu_1$, $\nu_2$, $R_{\rm br}$, and the flux normalization
are taken as free parameters. The spatial distribution of CR sources
follows that of supernova remnants
\begin{equation}
  f(r,z) = \left(\frac{r}{r_\odot}\right)^{1.25} 
  {\rm exp}\left(-3.56\cdot\frac{r-r_\odot}{r_\odot}\right)
  {\rm exp}\left(-\frac{|z|}{z_s}\right)\,,
  \label{spatial_distribution}
\end{equation}
which was adjusted to be consistent with the Galactic diffuse $\gamma$-ray
emission \cite{2011ApJ...729..106T}. In the above equation, $r_\odot=8.5$
kpc and $z_s=0.2$ kpc.

The propagation of charged particles is characterized by a diffusion
process in the random magnetic field, experiencing collisions and
energy losses due to interactions with gas and fields of the Milky
Way, and probably be reaccelerated by random magnetohydrodynamic (MHD)
waves or advectively transported \cite{2007ARNPS..57..285S}.
The diffusion coefficient, usually assumed to be spatially independent,
can be parameterized as a power-law of rigidity,
$D=\beta D_0(R/R_0)^{\delta}$, where $\beta$ is the velocity of a
particle in unit of light speed. Some modifications of the diffusion
coefficient were proposed to better fit the data, which will be
described in detail below. The advective (or convective) transportation
\cite{1976ApJ...208..900J} is assumed to be along the perpendicular
direction of the Galactic plane, with a convection velocity linearly
increasing from the disk to halo,
$\mathbf{V}_c=\mathbf{z}\cdot\mathrm{d}V_c/\mathrm{d}z$, where
$\mathbf{z}$ is the position vector in the vertical direction.
The reacceleration is described by a diffusion in the momentum space
with a diffusion coefficient $D_{pp}$, which is anti-proportional
to the spatial diffusion coefficient \cite{1994ApJ...431..705S}.
The magnitude of reacceleration is usually characterized by the Alfven
velocity ($v_A$) of the MHD wave. Finally, CR particles are confined
in a cylindrical volume with a half-height $z_h$.

Low energy particles would be modulated by the solar magnetic field
associated with solar activities. We employ the force-field approximation
to describe the solar modulation effect on the particle spectrum
\cite{1968ApJ...154.1011G}. The modulation potential, $\Phi$, depends
on solar activities and changes with time. Since most of the TOA data
used in this work were taken during the same period, a single modulation
potential would be enough.

We discuss two kinds of models in this work: 1) the diffusion convection
model with a break of the rigidity-dependence of the diffusion coefficient
\cite{1998ApJ...509..212S}
\begin{equation}
D=\beta D_0 \left(\frac{R}{R_0}\right)^{\delta_0}\left[\frac{1+(R/R_0)
^{\delta-\delta_0}}{2}\right],
\end{equation}
which is denoted as DC2\footnote{In Ref. \cite{2017PhRvD..95h3007Y} the
DC2 model was defined as $D=\beta D_0 (R/R_0)^{\delta}$ for $R>R_0$ and
$D=\beta D_0$ for $R\leq R_0$. In this work we employ a smooth break and
relax the low rigidity slope parameter $\delta_0$.}, and 2) the diffusion
reacceleration model with an $\eta$-term of the velocity-dependence
\cite{2010APh....34..274D}
\begin{equation}
D=\beta^{\eta} D_0(R/R_0)^{\delta},
\end{equation}
which is denoted as DR2. These two models represent in general two classes
of propagation models usually discussed in literature. The modifications
of the diffusion coefficient, basically giving faster diffusion of low
energy particles than the usual form, could be motivated due to the
resonant scatterings of CRs off the plasma waves which result in
dissipations of such waves \cite{2006ApJ...642..902P}. The other models
such as the plain diffusion model, the diffusion convection model without
break of the diffusion coefficient, and the traditional version of the
diffusion reacceleration model with $\eta=1$, are special examples of
the above two. In Ref. \cite{2017PhRvD..95h3007Y} an additional diffusion
reacceleration convection model was also discussed. However, it was found
that such a model did not improve the fitting compared with the DR2 model,
but was difficult to get converged.

\section{ACE-CRIS data}

The ACE-CRIS data are extracted from the ACE Science 
Center\footnote{http://www.srl.caltech.edu/ACE/ASC/level2/lvl2DATA\_CRIS.html},
adopting the same observational period as that of AMS-02 (from May 19,
2011 to May 26, 2016). The systematical uncertainties of the flux
measurements come from the geometry factor ($2\%$), the scintillating 
optical fiber trajectory efficiency ($2\%$), and the correction for
spallation in the instrument ($\sim1\%-5\%$) \cite{2009ApJ...698.1666G}. 
The total uncertainties are obtained through quadratically adding the 
statistical ones and the systematical ones together 
\cite{2009ApJ...698.1666G}. Since the energy ranges of Boron and Carbon
nuclei are different, their fluxes are interpolated to common energy
grids as in Ref. \cite{2009ApJ...698.1666G}. The ACE-CRIS data about 
the B/C ratio, the Carbon and Oxygen fluxes are give in Tables 
\ref{table:ace_bc} and \ref{table:ace_co}.

\begin{table}[!htb]
\caption {B/C ratio observed by ACE-CRIS from May 19, 2011 to May 26, 2016.}
\begin{tabular}{cc}
\hline \hline
$E_k$   & Ratio \\
(GeV/n) &       \\
\hline
0.072 & $0.268\pm0.011$ \\
0.085 & $0.265\pm0.012$ \\
0.100 & $0.270\pm0.013$ \\
0.120 & $0.271\pm0.014$ \\
0.142 & $0.270\pm0.015$ \\
0.170 & $0.272\pm0.016$ \\
\hline
\hline
\end{tabular}
\label{table:ace_bc}
\end{table}

\begin{table}[!htb]
\caption {Carbon and Oxygen fluxes observed by ACE-CRIS from May 19, 2011 
to May 26, 2016.}
\begin{tabular}{cc|cc}
\hline \hline
$E_k$   & Carbon Flux  & $E_k$   & Oxygen Flux  \\
(GeV/n) & (m$^{-2}$s$^{-1}$sr$^{-1}$(GeV/n)$^{-1}$)  & (GeV/n) & (m$^{-2}$s$^{-1}$sr$^{-1}$(GeV/n)$^{-1}$)   \\
\hline
0.068 & $2.40\pm0.07$ & 0.080 & $2.64\pm0.08$ \\
0.092 & $2.90\pm0.09$ & 0.108 & $3.17\pm0.10$ \\
0.117 & $3.41\pm0.11$ & 0.138 & $3.64\pm0.12$ \\
0.139 & $3.67\pm0.13$ & 0.165 & $3.84\pm0.13$ \\
0.159 & $3.90\pm0.14$ & 0.188 & $4.05\pm0.16$ \\
0.177 & $4.07\pm0.17$ & 0.210 & $4.15\pm0.18$ \\
0.195 & $4.09\pm0.19$ & 0.231 & $4.08\pm0.20$ \\
\hline
\hline
\end{tabular}
\label{table:ace_co}
\end{table}

\section{Results}

We use the B/C ratio data measured by AMS-02 \cite{2016PhRvL.117w1102A},
and ACE-CRIS, and the Carbon flux by Voyager \cite{2016ApJ...831...18C}, 
AMS-02 \cite{2017PhRvL.119y1101A}, and ACE-CRIS in the fitting. 
The measurements of $^{10}$Be/$^9$Be ratio \cite{1988SSRv...46..205S,
1998ApJ...501L..59C,1999ICRC....3...41L,2001ApJ...563..768Y,
2004ApJ...611..892H}, although have relatively large uncertainties, 
are also included in order to have a loose constraint on the lifetime 
of CRs in the Milky Way. The $^{10}$Be/$^9$Be ratios were mostly measured 
by experiments decades ago. The corresponding modulation potential is 
adopted to be $\Phi-0.2$ GV according to the study of time variation 
of the modulation potential \cite{2017PhRvD..95h3007Y}.

\subsection{Propagation parameters}

\begin{figure}[!htb]
\includegraphics[width=0.48\textwidth]{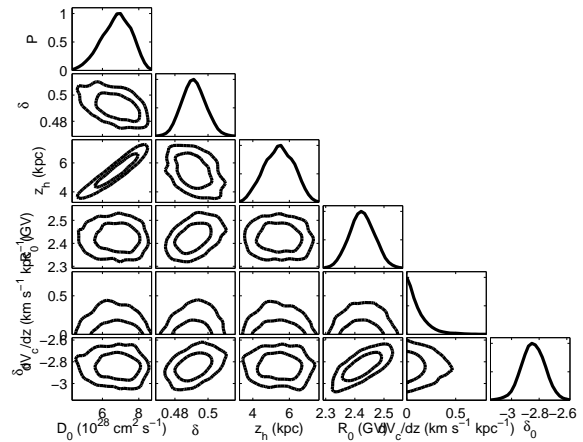}
\caption{Fitting 1-dimensional probability distributions (diagonal) 
and 2-dimensional credible regions (off-diagonal; $68\%$ and $95\%$ 
credible levels from inside to outside) of the propagation parameters
in the DC2 model. 
\label{fig:tri_DC2}}
\end{figure}

\begin{figure}[!htb]
\includegraphics[width=0.48\textwidth]{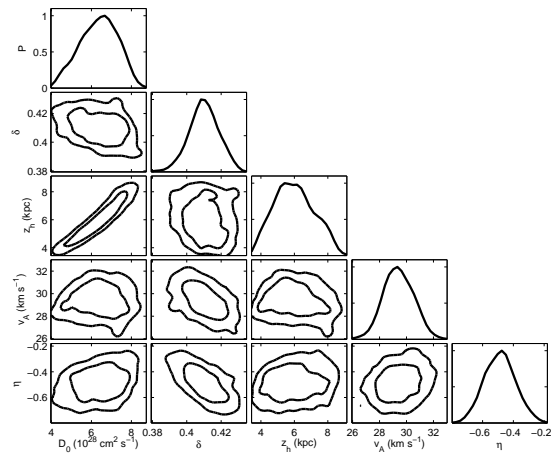}
\caption{Same as Fig.~\ref{fig:tri_DC2} but for the DR2 propagation
model.
\label{fig:tri_DR2}}
\end{figure}

The posterior mean values and the associated 68\% credible uncertainties
of the model parameters obtained from the fits to the B/C ratio and
Carbon flux data are tabulated in Table~\ref{table:para}. The corresponding
probability distributions of the propagation parameters are shown in
Figs.~\ref{fig:tri_DC2} and \ref{fig:tri_DR2}, for the DC2 and DR2 models, 
respectively. Comparing the minimum $\chi^2$ values of these two models, 
the DR2 model is significantly better than the DC2 model. The $\chi^2$ 
value for the DR2 model is smaller by 83.0 than that of DC2, with even 
less free parameters. Given the absolute value of $\chi^2$ and the number
of degree-of-freedom (dof), the DC2 model is marginally acceptable
(the $p$-value of the fit is about $0.056$).

\begin{table}
\caption {Posterior mean and $68\%$ credible uncertainties of the model
parameters}
\begin{tabular}{cccccccc}
\hline
\hline
                             & Unit                           & DC2               & DR2               \\
\hline
$D_0$                        & $(10^{28}{\rm cm^2s^{-1}})$    & $6.76 \pm 0.79$   & $6.46 \pm 0.88$   \\
$\delta_0$                   &                                & $-2.84 \pm 0.08$  & ---               \\
$\delta$                     &                                & $0.491 \pm 0.007$ & $0.410 \pm 0.009$ \\
$R_0$                        & $({\rm GV})$                   & $2.42 \pm 0.04$   & 4.0 (fixed)       \\
$z_h$                        & $({\rm kpc})$                  & $5.43 \pm 0.77$   & $6.11 \pm 1.14$   \\
$v_A$                        & $({\rm km\,s^{-1}})$           & ---               & $29.4 \pm 1.1$    \\
$dV_c/dz$                    & $({\rm km\,s^{-1}\,kpc^{-1}})$ & $<0.35^\dagger$   & ---               \\
$\eta$                       &                                & 1.0 (fixed)       & $-0.48 \pm 0.10$  \\
$X_{\rm C/H}^\ddagger$       & $(10^{-3})$                    & $3.32 \pm 0.02$   & $3.30 \pm 0.03$   \\
$\nu_1$                      &                                & $0.74 \pm 0.24$   & $0.64 \pm 0.12$   \\
$\nu_2$                      &                                & $2.37 \pm 0.01$   & $2.38 \pm 0.01$   \\
$R_{\rm br}$                 & $({\rm GV})$                   & $0.66 \pm 0.06$   & $1.51 \pm 0.07$   \\
$\Phi$                       & $({\rm GV})$                   & $0.694 \pm 0.011$ & $0.784 \pm 0.012$ \\
$\chi^2_{\rm min}/{\rm dof}$ &                                & 188.3/159         & 105.3/160         \\ \hline
\end{tabular}\\
$^\dagger$95\% upper limit.
$^\ddagger$Source abundance ratio of Carbon-to-Hydrogen. The normalization of the propagated proton flux at 100 GeV is $4.45 \times 10^{-9}~\mathrm{cm^{-2}s^{-1}sr^{-1}MeV^{-1}}$.
\label{table:para}
\end{table}

The convection velocity gradient, $dV_c/dz$, in the DC2 model is found 
to be very small. In such a case the DC2 model actually returns to the 
plain diffusion model. The parameter $v_A$ in the DR2 model is fitted
to be about $29.4$ km s$^{-1}$, which is smaller than that inferred
in the traditional diffusion reacceleration model configuration (i.e., 
$\eta=1$; hereafter DR) \cite{2011ApJ...729..106T,2015JCAP...09..049J,
2017PhRvD..95h3007Y,2018PhRvD..97b3015N}. This is due to the
$\beta^{\eta}$ term in the diffusion coefficient compensates to some
degree the reacceleration effect required to reproduce the bump of the 
B/C ratio (see below). The fit with DR2 model in Ref. 
\cite{2017PhRvD..95h3007Y} gives $v_A\approx18.4$, which is smaller
than that obtained here. This is perhaps due to the inclusion of
the Voyager data in the fit. 

In both models, the height of the propagation halos is constrained to
about $5\sim6$ kpc, which is similar to the canonical value of 4 kpc 
usually adopted. For comparison, the fit with the same DR2 model in
Ref. \cite{2017PhRvD..95h3007Y} gives $z_h=5.0\pm0.9$ kpc, which is
consistent with this work. The fits with the DR model configuration 
using different data sets gave $5.4\pm1.4$ kpc \cite{2011ApJ...729..106T}, 
$3.3\pm0.6$ \cite{2015JCAP...09..049J}, $5.9\pm1.1$ \cite{2017PhRvD..95h3007Y},
and $7.4\pm0.6$ \cite{2018PhRvD..97b3015N}. The value of $\delta$ is 
found to be about 0.5 (0.4) for the DC2 (DR2) model at high rigidities. 
Previous studies gave values from 0.3 to 0.5 for the DR model 
\cite{2011ApJ...729..106T,2015JCAP...09..049J,2017PhRvD..95h3007Y,
2018PhRvD..97b3015N}, and about $0.6$ for the convection model 
\cite{2017PhRvD..95h3007Y}, respectively.

\begin{figure}[!htb]
\includegraphics[width=0.48\textwidth]{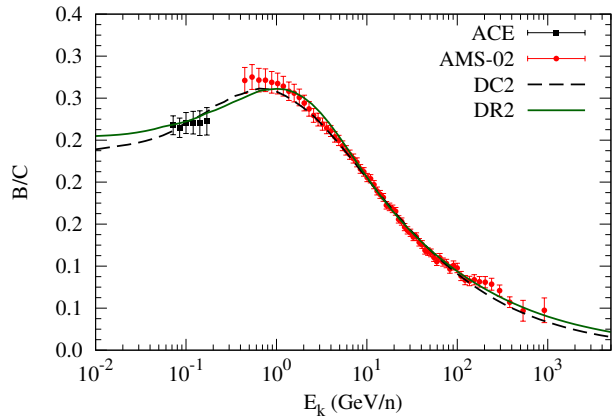}
\caption{Comparison between the best-fit model results of the B/C ratios
and the observational data. The dashed line is for the DC2 model, and 
the solid line is for the DR2 model.
\label{fig:BC}}
\end{figure}

\begin{figure}[!htb]
\includegraphics[width=0.48\textwidth]{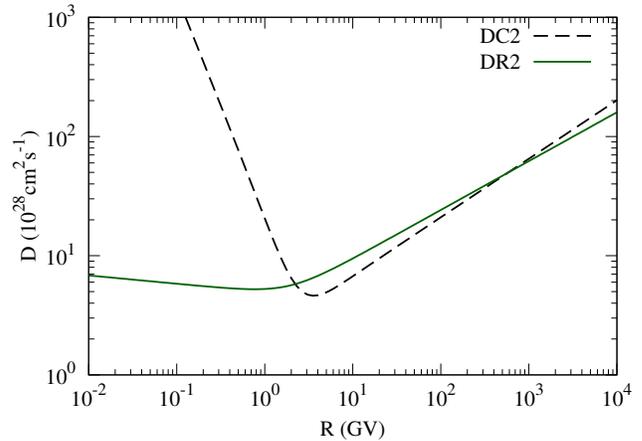}
\caption{Fitting diffusion coefficients of the DC2 (dashed) and DR2 (solid)
models. The results are calculated assuming $Z/A=1/2$.
\label{fig:Dxx}}
\end{figure}

Fig.~\ref{fig:BC} shows the best-fit results of the B/C ratios of the 
two models, compared with the data. We find that the DR2 model naturally 
gives a smooth bump of the B/C ratio, due to the reacceleration effect 
of low energy particles. The convection model is in general difficult 
to give such a bump \cite{2017PhRvD..95h3007Y}. In this work a bump is 
produced in the DC2 model due to the assumed spectral break of the 
diffusion coefficient. The low energy spectral index of the diffusion 
coefficient, $\delta_0$, is fitted to be about $-2.84$. This result 
indicates that low energy particles diffuse even faster than intermediate 
energy ones, resulting in a suppression of the low energy B/C and hence 
giving a bump. Nevertheless, the $\beta^{\eta}$ term in the DR2 model 
also suggests such a behavior of the diffusion coefficient at low energies. 
As a comparison, we show in Fig.~\ref{fig:Dxx} the diffusion coefficients 
in both models. We can see that for rigidities higher than a few GV,
the diffusion coefficients of these two models are similar with each
other. At low energies they differ from each other significantly.
Since these two models give similar B/C ratio, the difference of the
diffusion coefficients should be compensated by the effect of the 
reacceleration assumed in the DR2 model.

\subsection{Injection parameters}

We now turn to study the source injection spectra of CRs.
To properly take into account the uncertainties of the propagation
parameters, we follow Ref. \cite{2018JCAP...06..024C} to include
a prior term of the propagation parameters as
\begin{equation}
\chi^2_{\Sigma}=\left(\boldsymbol{\theta}-\bar{\boldsymbol{\theta}}\right)~
\Sigma^{-1}~\left(\boldsymbol{\theta}-\bar{\boldsymbol{\theta}}\right)^{T},
\end{equation}
where $\boldsymbol{\theta}$ represents the propagation parameter set,
$\bar{\boldsymbol{\theta}}$ is the vector of the mean propagation 
parameters given in Table \ref{table:para}, and $\Sigma$ is the 
covariance matrix of these parameters which has been derived in the
previous subsection.
We re-fit the fluxes of protons, Helium, Carbon, and Oxygen to 
derive the injection spectral parameters of these species. 
Slightly different from Eq.~(\ref{injection}), we introduce an 
additional spectral break at $R_{\rm br,2}$, which is around a few 
hundred GV, to describe the spectral hardenings. Besides the Voyager 
and AMS-02 data, for protons and Helium nuclei, the CREAM I+III 
data at higher energies are also used \cite{2017ApJ...839....5Y}. 
For Carbon and Oxygen nuclei, the ACE data given in Table 
\ref{table:ace_co} are used. Since the observed Helium, Carbon, 
and Oxygen spectra by AMS-02 show similarities among each other 
\cite{2017PhRvL.119y1101A}, we first do the fit with all these three 
compositions together. The injection parameters of them are assumed to 
be the same, with only differences of the normalizations. However, we 
find that in both DC2 and DR2 models, the goodness-of-fit is very poor. 
The propagated spectra of these nuclei differ mainly at low energies, 
perhaps due to different cooling rates of them in the interstellar 
medium. This result suggests that the injection spectra of Helium, 
Carbon, and Oxygen might be different. The similar result has also 
been obtained in a recent work \cite{2018ApJ...858...61B}.

\begin{figure*}[!htb]
\includegraphics[width=0.48\textwidth]{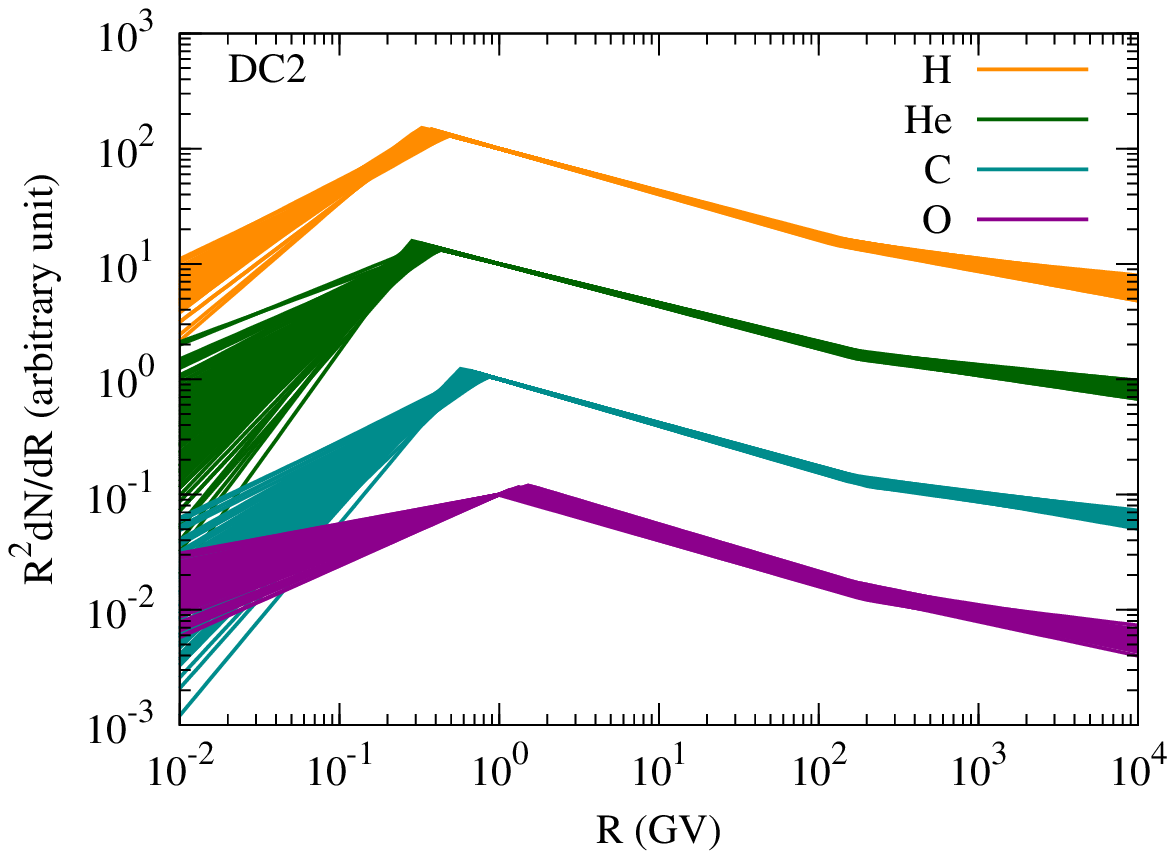}
\includegraphics[width=0.48\textwidth]{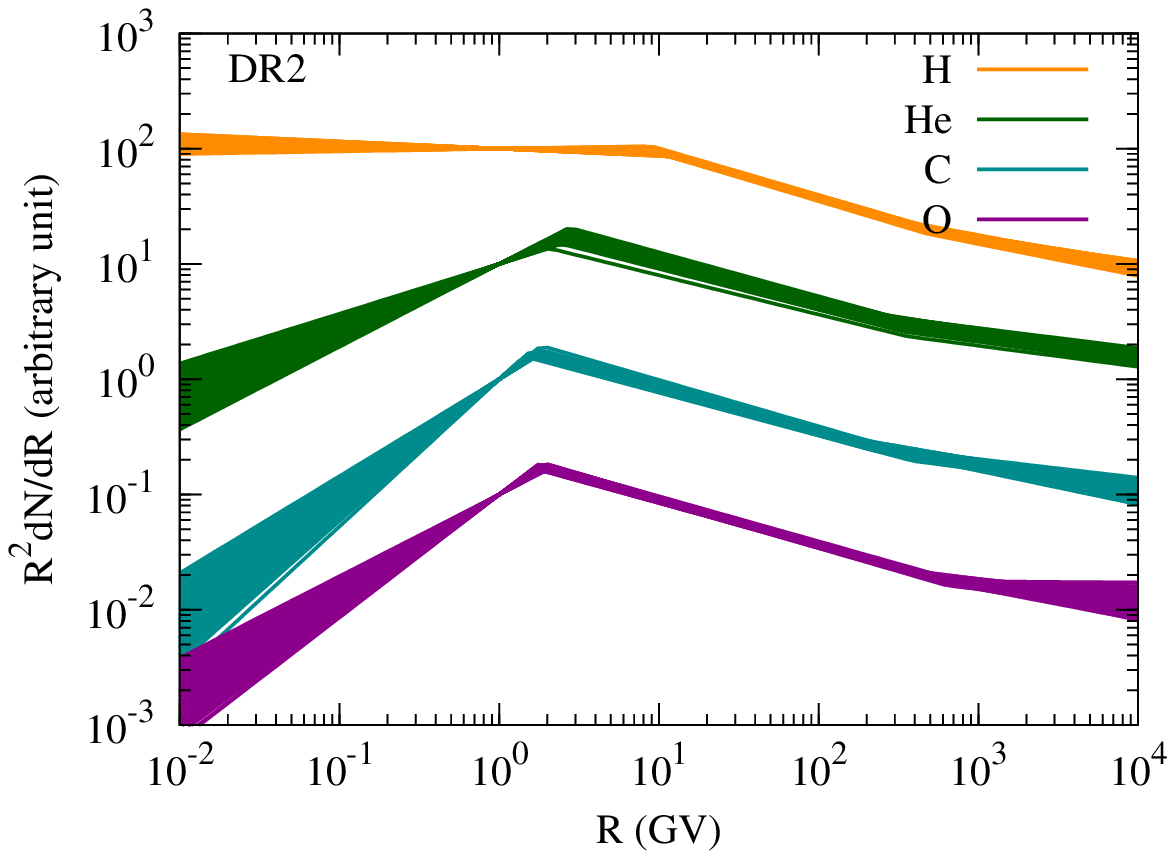}
\caption{Injection spectra for different nuclear compositions,
for the DC2 (left) and DR2 (right) propagation models. Bands show
the results with the $95\%$ ranges of the spectral parameters.
\label{fig:inj}}
\end{figure*}

\begin{figure*}[!htb]
\includegraphics[width=0.48\textwidth]{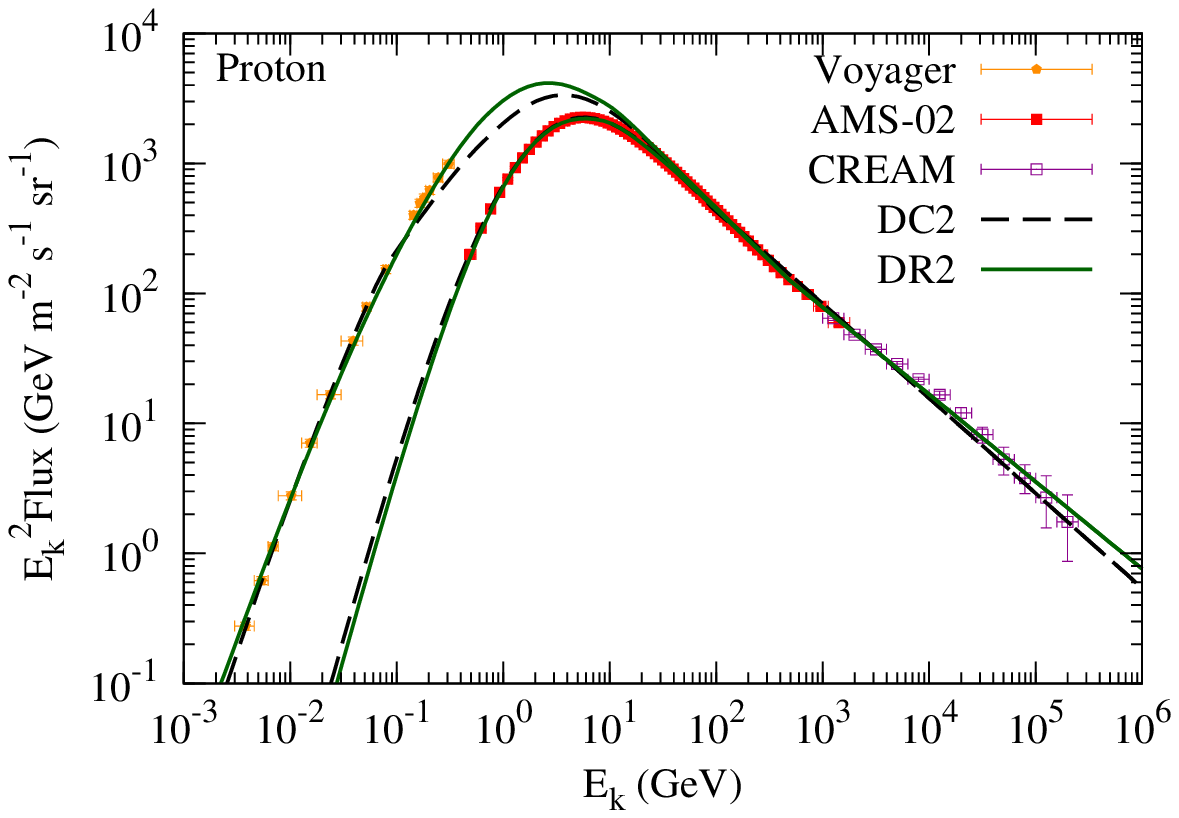}
\includegraphics[width=0.48\textwidth]{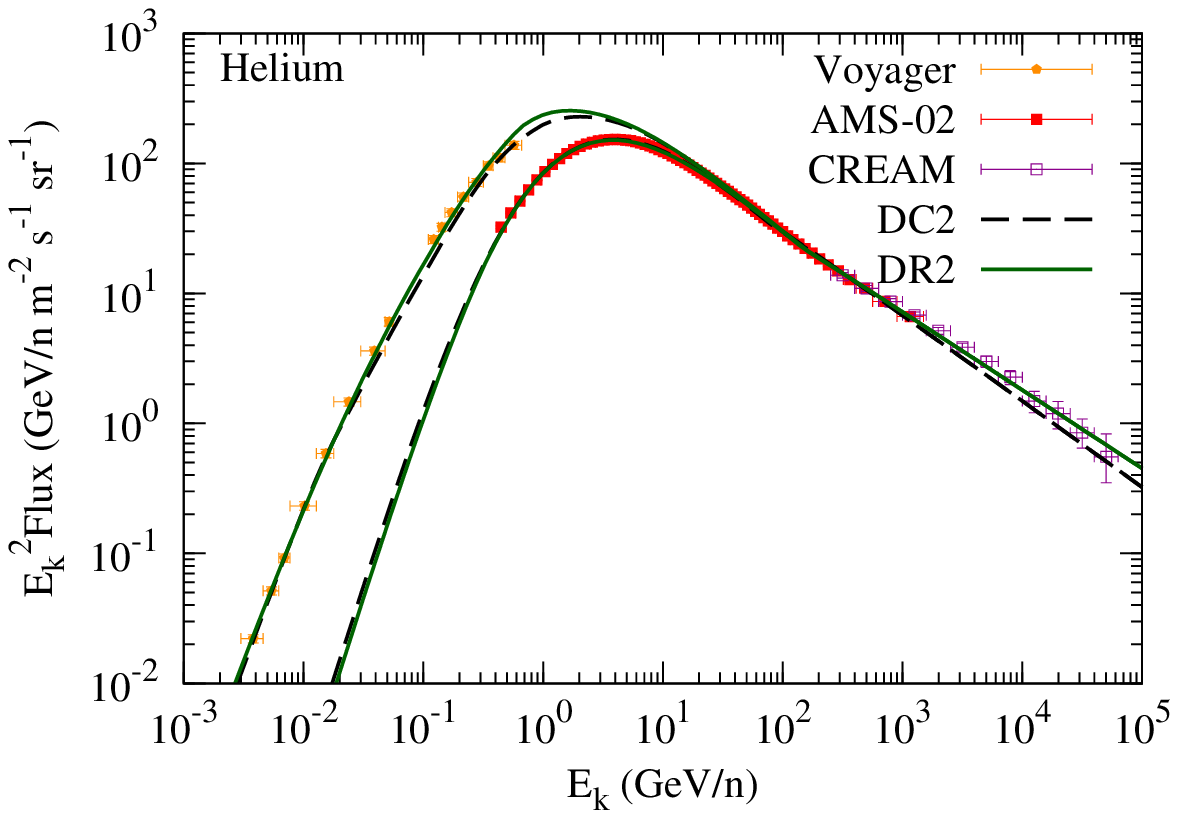}
\includegraphics[width=0.48\textwidth]{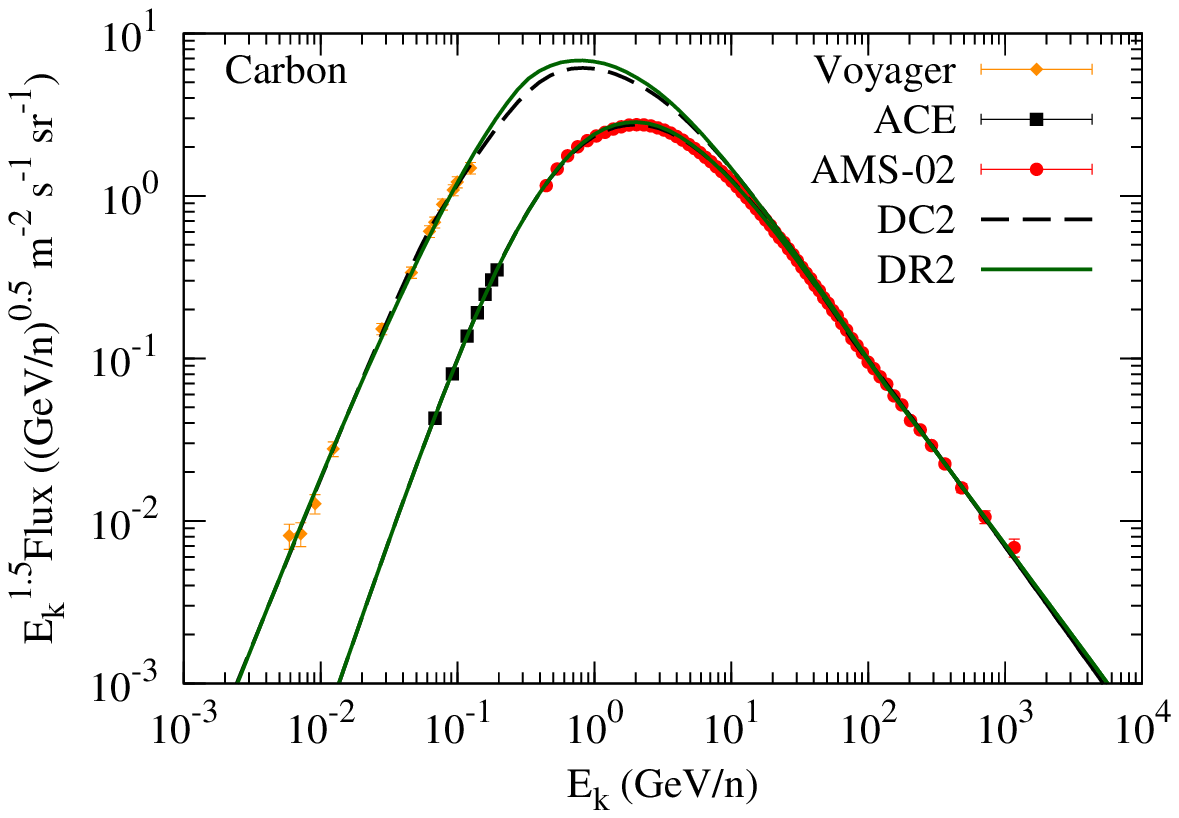}
\includegraphics[width=0.48\textwidth]{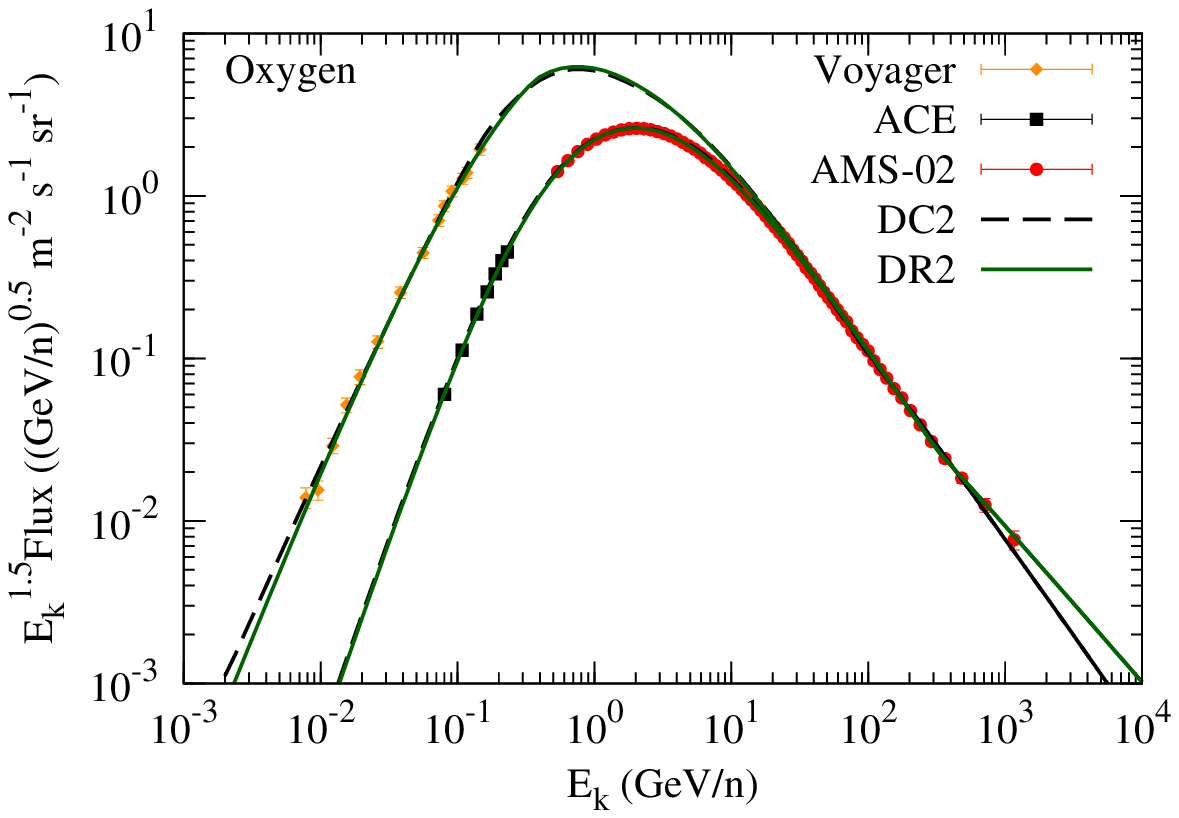}
\caption{Comparison between the best-fit model results of the proton
(left), Helium (middle), and Oxygen (right), and the observational data. 
The dashed lines are for the DC2 model, and the solid lines are for the DR2 
model. In each panel the upper lines are fluxes in the local interstellar
environment and the lower ones are those after the solar modulation.
\label{fig:pheco}}
\end{figure*}

\begin{table*}[!htb]
\caption {Fitting parameters of the injection spectra.}
\begin{tabular}{cccccc}
\hline \hline
    &  Para. &  H  &  He  &  C  &  O \\
\hline
DC2 & $X^\dagger$ ($10^{-3}$) & --- & $106.3\pm1.5$ & $3.77\pm0.07$ & $4.76\pm0.12$ \\
    & $\nu_1$ & $1.138\pm0.097$ & $0.857\pm0.323$ & $0.971\pm0.206$ & $1.580\pm0.092$ \\
    & $\nu_2$ & $2.379\pm0.008$ & $2.350\pm0.008$ & $2.390\pm0.007$ & $2.406\pm0.008$ \\
    & $\nu_3$ & $2.221\pm0.021$ & $2.167\pm0.017$ & $2.200\pm0.023$ & $2.230\pm0.034$ \\
    & $R_{\rm br}$ (GV) & $0.40\pm0.03$ & $0.35\pm0.03$ & $0.73\pm0.06$ & $1.25\pm0.12$ \\
    & $R_{\rm br,2}$ (GV) & $136.4\pm27.7$ & $184.0\pm20.4$ & $173.1\pm19.9$ & $203.5\pm33.0$ \\
    & $\Phi$ (GV) & $0.480\pm0.012$ & $0.613\pm0.016$ & $0.727\pm0.013$ & $0.733\pm0.014$ \\
    & $\chi^{2\ddagger}$/dof & 591.0/92 & 205.2/89 & 24.6/81 & 25.4/81 \\
\hline
DR2 & $X^\dagger$ ($10^{-3}$) & --- & $100.5\pm1.5$ & $3.59\pm0.10$ & $4.84\pm0.24$ \\
    & $\nu_1$ & $2.028\pm0.020$ & $1.436\pm0.071$ & $0.922\pm0.104$ & $1.134\pm0.079$ \\
    & $\nu_2$ & $2.405\pm0.007$ & $2.359\pm0.011$ & $2.383\pm0.007$ & $2.398\pm0.006$ \\
    & $\nu_3$ & $2.245\pm0.021$ & $2.193\pm0.020$ & $2.233\pm0.038$ & $2.173\pm0.068$ \\
    & $R_{\rm br}$ (GV) & $10.35\pm0.92$ & $2.56\pm0.15$ & $1.66\pm0.09$ & $1.92\pm0.10$ \\
    & $R_{\rm br,2}$ (GV) & $511.8\pm86.2$ & $302.9\pm36.1$ & $273.5\pm61.6$ & $488.4\pm85.2$ \\
    & $\Phi$ (GV) & $0.673\pm0.015$ & $0.728\pm0.015$ & $0.790\pm0.014$ & $0.772\pm0.012$ \\
    & $\chi^{2\ddagger}$/dof & 87.0/92 & 85.8/89 & 33.0/81 & 19.2/81 \\
\hline
\hline
\end{tabular}\\
$^\dagger$Source abundance relative to Hydrogen. 
$^\ddagger$$\chi^2_{\Sigma}$ has been included.
\label{table:inj}
\end{table*}

To further study the injection spectra of individual composition,
we fit the fluxes of protons, Helium, Carbon, and Oxygen nuclei
separately. The best-fitting injection spectra (with arbitrary
normalizations) are shown in Fig.~\ref{fig:inj}. Table~\ref{table:inj} 
lists the best-fitting $\chi^2$ values over the numbers of dof of 
all the fits. For the DC2 model, the injection spectra of protons,
Helium, and Carbon are similar. The low energy spectrum of Oxygen is
somewhat softer. However, we find that the fits of the proton and 
Helium spectra in the DC2 model are poor with quite large $\chi^2$ 
values. Therefore, the derived injection spectral parameters may 
have large bias and we mainly discuss the results of the DR2 model 
in the following. Compared with the DC2 model, the fitting results of 
the DR2 model are reasonable. These results disfavor the DC2 model. 
In the DR2 model, although the high-energy injection spectra of
different compositions are close to each other, the low-energy parts 
show a rough trend that lighter nuclei have softer spectra. The break 
energy of protons is also different from that of the other three. 
These differences may be related with the particle acceleration
process of these nuclei \cite{2017ApJ...844L...3Z}. 

Fig.~\ref{fig:pheco} shows the best-fit results of the proton (top-left),
Helium (top-right), Carbon (bottom-left), and Oxygen (bottom-right) fluxes 
for the DC2 and DR2 models. This plot illustrates that the DR2 model can
fit the data reasonably well, while the DC2 model matches with the 
low-energy data of Voyager and AMS-02 poorly for protons and Helium nuclei. 
A more complicated form of the injection spectrum may be necessary for 
the DC2 model (see e.g., \cite{2016ApJ...831...18C}).

\section{Conclusion and discussion}

In this work we employ the precise measurements of the CR nuclei data
by Voyager, ACE, AMS-02, and CREAM to study the propagation and injection
parameters of CRs. We first derive the Boron-to-Carbon ratios, and the 
Carbon and Oxygen fluxes recorded by the ACE-CRIS, during the same period 
of the AMS-02 results. The TOA data taken within the same time interval
are expected to have the same solar modulation effect, and can thus 
reduce the uncertainties due to solar modulation. The B/C and Carbon
fluxes measured by AMS-02 and ACE at TOA and by Voyager at outside of
the solar system are then used to constrain the CR propagation parameters, 
for two typical propagation settings (DC2 and DR2). Based on the obtained 
propagation parameters, we further investigate the source injection 
spectra of different nuclei.

We find that the propagation model with reacceleration (DR2) can fit 
the data significantly better than the other one (DC2). Even we have
assumed a convective propagation effect in the DC2 model, the fitting 
results disfavor a significant role of the convection. Nevertheless, 
the diffusion coefficient of the DR2 model needs a phenomenological 
modification at low energies to make those particles diffuse faster. 
This effect is even more obvious for the DC2 model, in which a break 
of the diffusion coefficient is assumed. This behavior of the diffusion 
coefficient could be due to the resonant interactions between such low 
energy CRs and the MHD waves which lead to dissipations of the MHD waves 
\cite{2006ApJ...642..902P}. The slopes of the diffusion coefficients at 
high rigidities are about $0.4\sim0.5$, which lie in between the 
predictions from the Kolmogrov ($\delta=1/3$; \cite{1941DoSSR..30..301K}) 
and the Kraichnan ($\delta=1/2$; \cite{1965PhFl....8.1385K}) types of 
the interstellar turbulence. The height of the propagation halo is 
constrained to be about $5\sim6$ kpc in both models.

The injection spectra of different nuclei are found to be different.
Even for Helium, Carbon, and Oxygen nuclei which have the same 
mass-to-charge ratios, their injection spectra are different, mainly
at low energies. The fits to the proton and Helium spectra in 
the DC2 model are poor, which again favors the DR2 model of the CR 
propagation. For the favored DR2 model, the injection spectra of 
lighter nuclei are found to be softer than that of heavier nuclei. 
The break energies also differ from each other. These results are 
expected to be useful for understanding the CR acceleration process, 
and may need further careful studies. 

The inclusion of the Voyager data in the fits can effectively break the 
degeneracy between the solar modulation model and the injection/propagation 
parameters. We find that for all the nuclei (except for protons and 
Helium in the DC2 model which give poor fits to the data), the force-field 
modulation potential is constrained to be about $0.7\sim0.8$ GV for the 
AMS-02 data taking time (May 19, 2011 to May 26, 2016). In this case we 
expect that the derived injection parameters are more directly relevant 
to the acceleration process of CR sources.

\acknowledgments
The author acknowledges the ACE-CRIS instrument team and the ACE Science 
Center for providing the ACE data. The author thanks Xiao-Jun Bi and 
Siming Liu for helpful discussion. This work is supported by the National 
Key R\&D Program of China (No. 2016YFA0400204), the National Natural 
Science Foundation of China (Nos. 11722328, U1738205, 11851305), and the 
100 Talents program of Chinese Academy of Sciences.


\begin{thebibliography}{43}

\bibitem{1990ApJ...349..625S}
S.~P. Swordy, D. Mueller, P.~Meyer, J.~L'Heureux, and J.~M. Grunsfeld, 
Astrophys. J. {\bf 349}, 625 (1990).

\bibitem{1991ApJ...374..356M}
D. Mueller, S. P. Swordy, P. Meyer, J. L’Heureux, and J. M.
Grunsfeld, Astrophys. J. 374, 356 (1991).

\bibitem{2001ApJ...555..585M}
D. Maurin, F. Donato, R. Taillet, and P. Salati, Astrophys. J.
{\bf 555}, 585 (2001).

\bibitem{2011ApJ...729..106T}
R. Trotta, G. Johannesson, I. V. Moskalenko, T. A. Porter, R. Ruiz de Austri, 
and A. W. Strong, Astrophys. J. {\bf 729}, 106 (2011).

\bibitem{2015JCAP...09..049J}
H.-B. Jin, Y.-L. Wu, and Y.-F. Zhou, J. Cosmol. Astropart. Phys.
{\bf 9}, 049 (2015).

\bibitem{2016ApJ...824...16J}
G. Johannesson, et al., Astrophys. J. {\bf 824}, 16 (2016).

\bibitem{2016PhRvD..94l3007F}
J. Feng, N. Tomassetti, and A. Oliva, Phys. Rev. D {\bf 94}, 123007 (2016).

\bibitem{2016PhRvD..94l3019K}
M. Korsmeier and A. Cuoco, Phys. Rev. D {\bf 94}, 123019 (2016).

\bibitem{2016PhRvL.117w1102A}
M. Aguilar, et al. (AMS collaboration), Phys. Rev. Lett. {\bf 117}, 
231102 (2016).

\bibitem{2017PhRvD..95h3007Y}
Q. Yuan, S.-J. Lin, K. Fang, and X.-J. Bi, Phys. Rev. D {\bf 95}, 083007 
(2017).

\bibitem{2018PhRvD..97b3015N}
J.-S. Niu and T. Li, Phys. Rev. D {\bf 97}, 023015 (2018).

\bibitem{2018JCAP...01..055R}
A. Reinert and M. W. Winkler, J. Cosmol. Astropart. Phys. {\bf 1},
055 (2018).

\bibitem{2017PhRvL.119y1101A}
M. Aguilar, et al. (AMS collaboration), Phys. Rev. Lett. {\bf 119}, 
251101 (2017).

\bibitem{2018PhRvL.120b1101A}
M. Aguilar, et al. (AMS collaboration), Phys. Rev. Lett. {\bf 120}, 
021101 (2018).

\bibitem{2012PhRvL.109f1101B}
P. Blasi, E. Amato, and P. D. Serpico, Phys. Rev. Lett. {\bf 109},
061101 (2012).

\bibitem{2012ApJ...752L..13T}
N. Tomassetti, Astrophys. J. Lett. {\bf 752}, L13 (2012).

\bibitem{2016ApJ...819...54G}
Y.-Q. Guo, Z. Tian, and C. Jin, Astrophys. J. {\bf 819}, 54 (2016).

\bibitem{2017PhRvL.119x1101G}
Y. Genolini, et al., Phys. Rev. Lett. {\bf 119}, 241101 (2017).

\bibitem{2018PhRvD..97f3008G}
Y.-Q. Guo and Q. Yuan, Phys. Rev. D {\bf 97}, 063008 (2018).

\bibitem{2018arXiv180203602L}
W. Liu, Y.-h. Yao, and Y.-Q. Guo, ArXiv e-prints (2018), 1802.03602.

\bibitem{2013Sci...341..150S}
E. C. Stone, A. C. Cummings, F. B. McDonald, B. C. Heikkila,
N. Lal, and W. R. Webber, Science {\bf 341}, 150 (2013).

\bibitem{2016ApJ...831...18C}
A. C. Cummings, E. C. Stone, B. C. Heikkila, N. Lal, W. R. Webber, 
G. Johannesson, I. V. Moskalenko, E. Orlando, and T. A. Porter, 
Astrophys. J. {\bf 831}, 18 (2016).

\bibitem{2010PhRvD..81b3516L}
J. Liu, Q. Yuan, X. J. Bi, H. Li, and X. M. Zhang, Phys. Rev. D
{\bf 81}, 023516 (2010).

\bibitem{2012PhRvD..85d3507L}
J. Liu, Q. Yuan, X.-J. Bi, H. Li, and X. Zhang, Phys. Rev. D {\bf 85},
043507 (2012).

\bibitem{1998ApJ...509..212S}
A. W. Strong and I. V. Moskalenko, Astrophys. J. {\bf 509}, 212 (1998).

\bibitem{1998ApJ...493..694M}
I. V. Moskalenko and A. W. Strong, Astrophys. J. {\bf 493}, 694 (1998).

\bibitem{2002PhRvD..66j3511L}
A. Lewis and S. Bridle, Phys. Rev. D {\bf 66}, 103511 (2002).

\bibitem{2007ARNPS..57..285S}
A. W. Strong, I. V. Moskalenko, and V. S. Ptuskin, Annual
Review of Nuclear and Particle Science {\bf 57}, 285 (2007).

\bibitem{1976ApJ...208..900J}
J. R. Jokipii, Astrophys. J. 208, 900 (1976).

\bibitem{1994ApJ...431..705S}
E. S. Seo and V. S. Ptuskin, Astrophys. J. {\bf 431}, 705 (1994).

\bibitem{1968ApJ...154.1011G}
L. J. Gleeson and W. I. Axford, Astrophys. J. {\bf 154}, 1011 (1968).

\bibitem{2010APh....34..274D}
G. di Bernardo, C. Evoli, D. Gaggero, D. Grasso, and L. Maccione, 
Astropart. Phys. {\bf 34}, 274 (2010).

\bibitem{2006ApJ...642..902P}
V. S. Ptuskin, I. V. Moskalenko, F. C. Jones, A. W. Strong,
and V. N. Zirakashvili, Astrophys. J. {\bf 642}, 902 (2006).

\bibitem{2009ApJ...698.1666G}
J. S. George, et al., Astrophys. J. {\bf 698}, 1666 (2009).

\bibitem{1988SSRv...46..205S}
J. A. Simpson and M. Garcia-Munoz, Space Sci. Rev. {\bf 46}, 205 (1988).

\bibitem{1998ApJ...501L..59C}
J. J. Connell, Astrophys. J. Lett. {\bf 501}, L59 (1998).

\bibitem{1999ICRC....3...41L}
A. Lukasiak, International Cosmic Ray Conference, {\bf 3}, 41 (1999).

\bibitem{2001ApJ...563..768Y}
N. E. Yanasak, et al., Astrophys. J. {\bf 563}, 768 (2001).

\bibitem{2004ApJ...611..892H}
T. Hams, et al., Astrophys. J. {\bf 611}, 892 (2004).

\bibitem{2018JCAP...06..024C}
M.-Y. Cui, X. Pan, Q. Yuan, Y.-Z. Fan and H.-S. Zong, 
J. Cosmol. Astropart. Phys. {\bf 6}, 024 (2018).

\bibitem{2017ApJ...839....5Y}
Y. S. Yoon, et al. (CREAM collaboration), Astrophys. J. {\bf 839}, 5 (2017).

\bibitem{2018ApJ...858...61B}
M. J. Boschini, et al., Astrophys. J. {\bf 858}, 61 (2018).

\bibitem{2017ApJ...844L...3Z}
Y. Zhang, S. Liu, and Q. Yuan, Astrophys. J. Lett. {\bf 844}, L3 (2017).

\bibitem{1941DoSSR..30..301K}
A. Kolmogorov, Akademiia Nauk SSSR Doklady {\bf 30}, 301 (1941).

\bibitem{1965PhFl....8.1385K}
R. H. Kraichnan, Physics of Fluids {\bf 8}, 1385 (1965).

\end{thebibliography}
\end{document}